\documentclass[final,1p,times]{elsarticle} 

\usepackage{graphicx}
\usepackage{amssymb} 
\usepackage{amsthm} 
\usepackage{lineno}
\usepackage{subfigure}

\journal{Nuclear Physics A} 

\begin{document}

\begin{frontmatter} 

\title{Influence of an inhomogeneous and expanding medium on signals of the QCD phase transition}

\author[auth1]{Marlene Nahrgang}

\author[auth2]{Christoph Herold}
\author[auth2]{Marcus Bleicher}
\address[auth1]{SUBATECH, UMR 6457, Universit\'e de Nantes, Ecole des Mines de Nantes,
IN2P3/CRNS. 4 rue Alfred Kastler, 44307 Nantes cedex 3, France}
\address[auth2]{Frankfurt Institute for Advanced Studies (FIAS), Ruth-Moufang-Str.~1, 60438 Frankfurt am Main, Germany}

\begin{abstract} 
According to a fluid dynamic expansion of the fireball we investigate how the inhomogeneity of the system influences the chiral phase transition of QCD. We compare the averaged values of the order parameter in equilibrium with that of a homogeneous system. If the temperature is averaged over a certain region of the fireball the corresponding correlation length does not diverge in an expansion with a critical point.
\end{abstract} 

\end{frontmatter} 


\section{Introduction}
The search for a possible critical point of QCD is one of the driving motivations for the low energy runs at RHIC/BNL and the upcoming facilities FAIR at GSI/Darmstadt and NICA in Dubna. The assumed existence of a critical point is based on two observations: at vanishing baryochemical potential highly sophisticated lattice QCD calculations revealed an analytic crossover  \cite{Aoki:2006we} and at high baryochemical potentials various approaches to effective models in mean-field and beyond find a first order phase transition, which then necessarily terminates in a critical point \cite{Scavenius:2000qd, Ratti:2005jh, arXiv:0704.3234,Skokov:2010wb}.

At both, the first order phase transition and the critical point, an enhancement of fluctuations is predicted to be seen in heavy-ion collisions. For the first order phase transition these are based on the growth of fluctuations due to the nonequilibrium effect of supercooling leading to nucleation and spinodal decomposition \cite{Csernai:1992tj,Mishustin:1998eq,Randrup:2010ax,Chomaz:2003dz}. Such fluctuations should be visible in studies of single event particle distributions. The critical point, however, is supposed to imprint its signals in event-by-event fluctuations of conserved quantities, such as net-charge or net-baryon number distributions \cite{Stephanov:1998dy,Stephanov:1999zu}. In thermodynamic systems the correlation length diverges. Systems created in a heavy-ion collision, however, are finite in size, highly dynamic and a temperature and baryochemical potential can at most be defined locally. The finite size of the system naturally limits the growth of the correlation length. The finite time, which the dynamic systems spends at a critical point, limits the growth even further because relaxation times also become infinite at the critical point. This phenomenon is called critical slowing down. 
Assuming a phenomenological time evolution of the correlation length with parameters from the $3$d Ising universality class it was found that the correlation length does not grow beyond $2-3$ fm \cite{Berdnikov:1999ph}. In all of these approaches the background medium is assumed to be homogeneous in its thermodynamic quantities, like the temperature. 

Recently, the model of nonequilibrium chiral fluid dynamics (N$\chi$FD) for the explicit propagation of fluctuations coupled to a dynamic expansion of a heavy-ion collision was developed \cite{Nahrgang:2011mg} showing that nonequilibrium effects can have an important influence on the evolution of the fluctuations of the order parameter of chiral symmetry \cite{Nahrgang:2011ll,arXiv:1105.1962}.
In this model the expansion of the fireball created in a heavy-ion collision is described by ideal fluid dynamics, where the fluid consists of quark and antiquarks with the local thermodynamic properties of the quark-meson model \cite{Scavenius:2000qd}. 

In this work we want to investigate the influence of the inhomogeneity of the fluid dynamic background on the sigma field and the correlation length. For this purpose we restrict ourself to a local equilibrium treatment of the sigma field, instead of an explicit propagation but work with concepts that have already been studied in N$\chi$FD. We refer to these references \cite{Nahrgang:2011mg, Nahrgang:2011ll,arXiv:1105.1962} for more information.

\section{Equilibrium chiral fluid dynamics}
For a fixed baryochemical potential $\mu_B$ the strength of the phase transition can be tuned via the quark-meson coupling $g$. At $\mu_B=0$ and for a realistic coupling of $g=3.3$ the transition is a crossover, for $g=3.63$ it is a critical point with $T_c\simeq140$~MeV and a first order phase transition with $T_c\simeq123$~MeV is found for $g=5.5$.

The thermodynamic potential of the quark-meson model in mean-field approximation is
\begin{equation}
 \Omega(T)=-\frac{T}{V}\ln Z=U\left(\sigma, \vec{\pi}\right)+\Omega_{q\bar q}(T)\, ,
\label{eq:lsm_omegaq0}
\end{equation}
with the chiral potential $U(\sigma,\vec\pi)$ and the quark contribution to fermionic one-loop level.
The equilibrium values for the chiral fields, $\sigma_{\rm eq}$ and $\vec\pi_{\rm eq}$, are obtained from minimizing the thermodynamic potential (\ref{eq:lsm_omegaq0}) with respect to $\sigma$ and $\vec\pi$. While $\sigma_{\rm eq}$ becomes finite at lower temperatures due to dynamic symmetry breaking, $\vec\pi$ vanishes for all temperatures. We do not continue to explicitly mention the pion field in the notation.
The minima of the thermodynamic potential $\Omega$ define the thermodynamically stable states of the matter. In this case the pressure and the energy density of the system are
\begin{equation}
  p(T)=-\Omega(T)|_{\sigma=\sigma_{\rm eq}}\, ,\quad\quad
 e(T)= T\left(\frac{\partial p(T)}{\partial T}\right)-p(T)\, .
\end{equation}
The fermions are described fluid dynamically by the relativistic equation of energy-momentum conservation
\begin{equation}
 \partial_\mu T^{\mu\nu} =0\; ,
\label{eq:fluidT}
\end{equation}
with the energy-momentum tensor of an ideal fluid.
Starting from initial conditions from UrQMD \cite{Petersen:2008dd} the fluid has an inhomogeneous temperature distribution, which is determined by the evolution of the energy density. Therefore, the equilibrium value of the sigma field is defined only locally in each cell of the fluid dynamic grid. Speaking of the temperature of the system thus implies some form of averaging over parts of the system. As a consequence, all quantities deduced from thermodynamic expressions are subject to the same averaging.

A prominent feature in a scenario with a first order phase transition is the discontinuity of the sigma field at the transition temperature, which will be studied in section \ref{sec:sec1}. In a critical point scenario we investigate the correlation of the fluctuations based on the definition via the curvature of the thermodynamic potential in section \ref{sec:sec2}.

\section{Discontinuity in a first order phase transition scenario}\label{sec:sec1}

\begin{figure}[htbp]
 \begin{center}
 \includegraphics[width=0.52\textwidth]{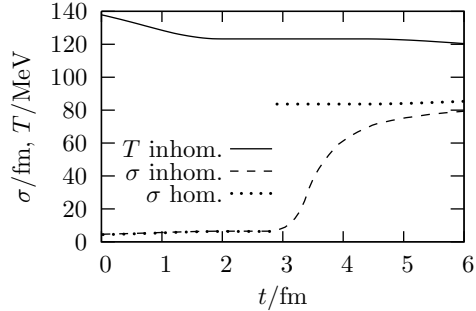}
 \end{center}
 \caption{Scenario with a first order phase transition. Time evolution of the averaged temperature (solid line) and the averaged sigma field (dashed line) in a fluid dynamically expanding system, where the local sigma field is in equilibrium with the fluid. Assuming that this temperature evolution was the global temperature of a homogeneous system, we show the global sigma field (dotted line) for comparison.}
 \label{fig:fo}
 \end{figure}

In figure \ref{fig:fo} we present the temperature and the sigma field averaged over an inner volume of about $30$~fm$^3$. As the system expands it dilutes and the energy density and the temperature become smaller until the coexistence region of the first order phase transition is reached. Here, the local  temperature stays constant. Since almost all of the volume has an energy density in the coexistence region, also the averaged temperature stays almost constant. The averaged sigma field increases during the transition. If we assumed a homogeneous temperature of the system with the same evolution as the averaged temperature in figure \ref{fig:fo} and assumed the corresponding equilibrium value of the sigma field for the entire system, we recover the discontinuity in the evolution of the sigma field. This discontinuity is washed out when averaging over an inhomogeneous system.

\section{Correlation length in a critical point scenario}\label{sec:sec2}
\begin{figure}[htbp]
 \begin{center}
 \subfigure{\includegraphics[width=0.47\textwidth]{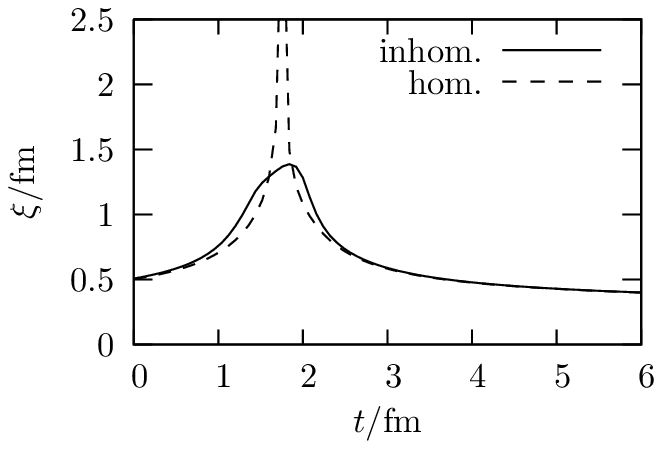}}
 \subfigure{\includegraphics[width=0.52\textwidth]{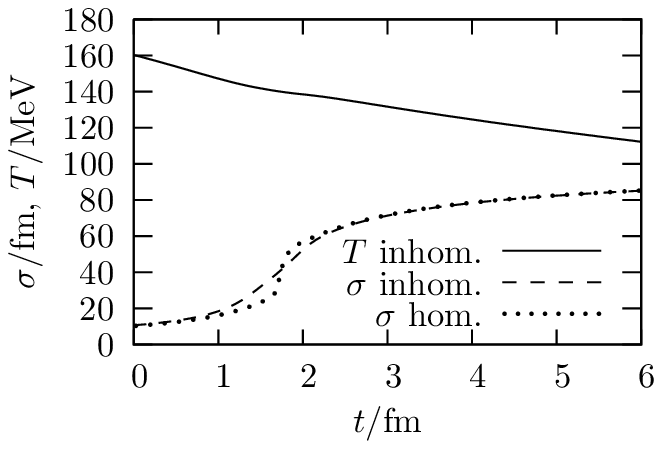}}
 \end{center}
 \caption{Scenario with a critical point. The curves in the right plot are the same as in figure \ref{fig:fo} but for a critical point. The left plot shows the averaged correlation length (solid line) compared to a correlation length corresponding to a homogeneous system temperature (dashed line).}
 \label{fig:cp}
 \end{figure}

In the right plot of figure \ref{fig:cp} the same curves are plotted as in figure \ref{fig:fo} but here for a scenario with a critical point. Again we find that in the transition region the evolution of the sigma field is washed out in an inhomogeneous system after averaging compared to an assumed homogeneous temperature evolution.
In the left plot we show the influence of the inhomogeneity on the correlation length, which is locally evaluated from 
\begin{equation}
 \frac{1}{\xi^2}=\frac{\partial^2\Omega(T)}{\partial\sigma^2}\bigg|_{\sigma=\sigma_{\rm eq}}\, ,
\end{equation}
We observe that due to averaging the correlation length in an inhomogeneous system does not diverge at the averaged critical temperature of the system. It remains finite and grows up to $1.5$~fm. When we assume again that the temperature evolution was a homogeneous temperature of the system we can calculate a single correlation length, which would correspond to the system's correlation length. It is shown in figure \ref{fig:cp} for comparison.

In finite systems one speaks about a critical region around the critical point, where the correlation length is still significantly enhanced but does not diverge. For an inhomogeneous system we find the same situation when averaging over parts of the system.

\section{Summary}

We showed that the inhomogeneity of a system created in a heavy-ion collision smears out the evolution of the order parameter of chiral symmetry, the sigma field, in the transition region. The effect on the correlation length is more dramatic, as its divergence at the critical temperature in a homogeneous system is reduced to an increase up to $1.5$~fm. In the investigated scenario, however, fluctuations of the sigma field are not directly studied. It needs to be said that the growth of the correlation length is a very dynamic process and should be studied in a fully dynamic setup, what is currently worked on within N$\chi$FD.

\vskip2em
This work was supported by the Hessian Excellence Inititive LOEWE through the Helmholtz International Center for FAIR. We are grateful to the Center for Scientific Computing (CSC) at Frankfurt for providing the computing resources.


\end{document}